\documentclass{elsarticle}

\usepackage{amsmath,amstext,amssymb}
\usepackage{subfigure,color}
\usepackage{psfrag,subfigure,color}
\usepackage{psfrag,subfigure,xcolor,color}
\usepackage{graphicx}
\usepackage{calc}

\def\ie{{i.\,e.\ }}
\def\eg{{e.\,g.\ }}

\def\k{\kappa_5}

\def\del{\partial}

\def\ee{{\mathrm e}}
\def\ii{{\mathrm i}}

\newcommand{\dd}{\mathrm{d}}

\begin{document}


\title{Non-universal shear viscosity from Einstein gravity}
\date{\today}

\cortext[cor1]{Corresponding Author}

\author[MPI]{Johanna Erdmenger}
\ead{jke@mppmu.mpg.de}
\author[MPI]{Patrick Kerner\corref{cor1}}
\ead{pkerner@mppmu.mpg.de}
\author[MPI]{Hansj\"org Zeller}
\ead{zeller@mppmu.mpg.de}

\address[MPI]{Max-Planck-Institut f\"ur Physik (Werner-Heisenberg-Institut)\\
  F\"ohringer Ring 6, 80805 M\"unchen, Germany}


\begin{abstract}
A very famous result of gauge/gravity duality is the universality of the ratio of shear viscosity to entropy density in every field theory holographically dual to classical, two-derivative (Einstein) gravity. We present a way to obtain deviation from this universality by breaking the rotational symmetry spontaneously. In anisotropic fluids additional shear modes exist and their corresponding shear viscosities may be non-universal. We confirm this by explicitly calculating the shear viscosities in a transversely isotropic background, a p-wave superfluid, and study its critical behavior. This is a first decisive step towards further applications of gauge/gravity duality to physical systems.
\end{abstract}


\maketitle

Gauge/gravity duality \cite{Aharony:1999ti} provides a novel
method for studying strongly-coupled systems at finite temperature and
densities. As such it is expected to have useful applications to the quark-gluon
plasma as well as to condensed matter physics. The duality also allows
to calculate physical observables in the real-time formalism, \eg
retarded correlation functions \cite{Son:2002sd,Herzog:2002pc} and
hydrodynamic transport coefficients \cite{Son:2007vk}. The most
important result of these studies is the universality of the ratio of
the shear viscosity $\eta$ to the entropy density $s$ in every field theory holographically dual to 
classical, two-derivative (Einstein) gravity theory
\cite{Kovtun:2004de,Buchel:2003tz,Benincasa:2006fu,Iqbal:2008by}. The
universal number $\eta/s=1/4\pi$ (in natural units) fits surprisingly
well the measured quantities in strongly-coupled real-world systems such as the
quark-gluon plasma and cold fermions at unitarity, see \eg
\cite{Schafer:2009dj}. The classical, two-derivative gravity theory is mapped to a gauge theory with large rank of the gauge
group (large $N_c$) and to a large 't Hooft coupling $\lambda$.

Since the ratio of shear viscosity to entropy density usually depends on the temperature and is non-universal in real-world systems, it is important to find deviations from the universal behavior in order to make further contact with real-world systems. In this letter, we present a deviation from the universal ratio at leading order in $N_c$  and $\lambda$. So far a deviation from the universality only occurs in subleading corrections of $N_c$ and/or $\lambda$, see \eg \cite{Buchel:2008wy, Buchel:2008vz}.  

The universality of $\eta/s$ strongly depends on the shear mode transforming as a helicity two state under the rotational symmetry. Since it is the only helicity two state in the system, the shear mode decouples from the other modes and it can effectively be written as a minimally coupled scalar. The universality of $\eta/s$ is then related to the universality of the coupling of this minimally coupled scalar. The coupling is always given by the gravitational coupling constant since the shear mode is a graviton \cite{Iqbal:2008by}. We circumvent this important assumption that the shear mode is a helicity two state by breaking the rotational symmetry spontaneously. A fluid with spontaneously broken rotational symmetry is anisotropic and more than one shear mode exists. Since some of these modes do not transform as helicity two states, the corresponding shear viscosities can be non-universal \cite{Landsteiner:2007bd, Natsuume:2010ky}.

 The viscosity, including shear and bulk viscosity, in anisotropic fluids is described by a rank four tensor with in general 21 independent coefficients \cite{Landau:1959te}. In the following we study systems with p-wave symmetry (transversely isotropic). This reduces the independent coefficients in the viscosity tensor to five. Two of these independent coefficients  are shear viscosities \cite{LESLIE01011966}. We  explicitly calculate these two shear viscosities by using the recipe for the holographic calculation of the retarded correlators  \cite{Son:2002sd,Herzog:2002pc} and Kubo formulae \cite{Sarman:1993sm} which are given by
\begin{equation}
\label{eq:Kubo}
\eta_{i}=-\lim_{\omega\to 0}\frac{1}{\omega}\mathrm{Im}\; G^R_{i,i}(\omega,0)\quad\text{with}\quad i\in\{xy,xz,yz\}\,,
\end{equation}
where the $x$-axis is the preferred direction and $\eta_{xy}(=\eta_{xz})$, $\eta_{yz}$ are the two independent shear viscosities. The retarded correlation function of the energy-momentum tensor $T_{\mu\nu}$ is defined by
\begin{equation}
\label{eq:defcorrelator}
G^R_{ij,kl}(\omega,0)=-\ii\int\!\dd t\,\dd x\,\ee^{\ii\omega t}\theta(t)\langle[T_{ij}(t,x),T_{kl}(0,0)]\rangle\,.
\end{equation}
We present more details on anisotropic shear viscosities in the
appendix below.

In the holographic context, the spontaneous breaking of symmetries by black holes developing hair was
first achieved in \cite{Gubser:2008px} and later used to construct
holographic superconductors/superfluids by breaking an abelian
symmetry \cite{Hartnoll:2008vx}. Along this line also p-wave
superconductors/superfluids have been constructed \cite{Gubser:2008wv} and
gave rise to the first string theory embeddings of holographic
superconductors/superfluids
\cite{Ammon:2008fc,Basu:2008bh,Ammon:2009fe}. In p-wave
superconductors/superfluids, also the spatial rotational symmetry is
spontaneously broken in addition to the internal abelian symmetry. In order to obtain the effects of spontaneous rotational symmetry breaking for the energy-momentum tensor, which determines the hydrodynamics of the systems, we have to take  the back-reaction of the superfluid into account. This was obtained \eg in \cite{Ammon:2009xh}. On the gravity side, the p-wave superfluid state corresponds to an asymptotically AdS black hole which carries vector hair.

In order to construct p-wave superfluid states, we consider $SU(2)$ Einstein-Yang-Mills theory in $(4+1)$-dimensional asymptotically AdS space as in \cite{Ammon:2009xh}. The action is
\begin{equation}
\label{eq:action}
S = \frac{1}{2\k^2}  \int\!\dd^5x\sqrt{-g} \: \left [\left( R -\Lambda\right) - \frac{\alpha^2}{2} \, F^a_{\mu\nu} F^{a\mu \nu} \right] \,, 
\end{equation}
where $\k$ is the five-dimensional gravitational constant, $\Lambda = - \frac{12}{L^2}$ is the cosmological constant, with $L$ being the AdS radius, and $\alpha=\kappa_5/\hat{g}$ the ratio of the gravitational constant $\kappa_5$ to the Yang-Mills coupling constant $\hat g$. The $SU(2)$ field strength $F^a_{\mu\nu}$ is
\begin{equation}
F^a_{\mu\nu}=\del_\mu A^a_\nu -\del_\nu A^a_\mu + \epsilon^{abc}A^b_\mu A^c_\nu \,,
\end{equation}
where $\mu, \nu = \{t,r,x,y,z\}$, with $r$ being the AdS radial coordinate, and $\epsilon^{abc}$ is the totally antisymmetric tensor with $\epsilon^{123}=+1$. The $A^a_\mu$ are the components of the matrix-valued gauge field, $A=A^a_\mu\tau^a dx^\mu$,  where the $\tau^a$ are the $SU(2)$ generators, which are related to the Pauli matrices by $\tau^a=\sigma^a/2\ii$.

Following ref.~\cite{Gubser:2008wv}, to construct charged black hole solutions with vector hair we choose a gauge field ansatz
\begin{equation}
\label{eq:gaugefieldansatz}
A=\phi(r)\tau^3\dd t+w(r)\tau^1\dd x\,.
\end{equation}
The motivation for this ansatz is as follows. In the field theory we
introduce a chemical potential for the $U(1)$ symmetry generated by $\tau^3$.
We denote this $U(1)$ as $U(1)_3$. The gravity field dual to the $U(1)_3$
density is $A^3_t$, hence we include $A^3_t(r) \equiv \phi(r)$ in our ansatz.
We allow for states with a nonzero $\langle J^x_1\rangle$, so in
addition we introduce $A^1_x(r) \equiv w(r)$. With this ansatz for the gauge
field, the Yang-Mills stress-energy tensor is
diagonal. Solutions with nonzero $w(r)$ preserve only an $SO(2)$ subgroup of the $SO(3)$ rotational symmetry, so
our metric ansatz respects only $SO(2)$. Furthermore, given that the
Yang-Mills stress-energy tensor is diagonal, a diagonal metric is consistent.  Our metric ansatz is  
\begin{equation}\label{eq:metricansatz}
\begin{split}
\dd s^2 = &-N(r)\sigma(r)^2\dd t^2 + \frac{1}{N(r)}\dd r^2 +r^2 f(r)^{-4}\dd x^2\\
& + r^2f(r)^2\left(\dd y^2 + \dd z^2\right)\,,
\end{split}
\end{equation}
with $N(r)=-\frac{2m(r)}{r^2}+\frac{r^2}{L^2}$. For our black hole solutions we denote the position of the horizon as $r_h$. The AdS boundary is at $r\rightarrow\infty$. 

Inserting our ansatz into the Einstein and Yang-Mills equations yields five equations of motion for  $m(r),\,\sigma(r),\,f(r),\,\phi(r),\,w(r)$ and one constraint equation from the $rr$ component of the Einstein equations. The dynamical equations can be found in  \cite{Ammon:2009xh}. Using scale transformations, we can set the boundary values of both $\sigma(r)$ and $f(r)$ to one, so that the metric will be asymptotically AdS.

A known analytic solution of the equations of motion is an asymptotically AdS Reissner-Nordstr\"om black hole, which has $\phi(r)=\mu - q/r^2$, $w(r)=0$, $\sigma(r)=f(r)=1$, and $N(r)= \left(r^2 - \frac{2m_0}{r^2} + \frac{2\alpha^2 q^2}{3 r^4}\right)$, where $m_0=\frac{r_h^4}{2}+\frac{\alpha^2 q^2}{3r_h^2}$ and $q= \mu r^2_h$. Here $\mu$ is the value of $\phi(r)$ at the boundary, which in CFT terms is the $U(1)_3$ chemical potential. Since $w=0$, \ie  $\langle J^x_1\rangle=0$, this solution corresponds to the normal phase of the system. To find solutions with nonzero $w(r)$ we resort to numerics. We solve the equations of motion using a shooting method (see \cite{Ammon:2009xh} for details).

We calculate the shear viscosity in this background via the Kubo formulae by considering two point functions of the energy-momentum tensor and the currents \footnote{The two point functions of the currents have to be considered in addition to the two point functions of the energy-momentum tensor since there are interactions between the energy-momentum tensor and the currents.}. As described in \cite{Son:2002sd,Herzog:2002pc,Kaminski:2009dh}, these two point functions  are determined by considering fluctuations of the metric $h_{\mu\nu}$ and the gauge fields $a_\mu^a$ about this background.  For these fluctuations we choose the ansatz
\begin{equation}
\label{eq:ansatzfluc}
\begin{split}
h_{\mu\nu}(t,\vec{x},r)&=h_{\mu\nu}(r)\ee^{-\ii\omega t+\ii \vec{k}\cdot\vec{x}}\,,\\
a^a_\mu(t,\vec{x},r)&=a^a_\mu(r)\ee^{-\ii\omega t+\ii \vec{k}\cdot\vec{x}}\,.
\end{split}
\end{equation}
Taking  the gauge freedom into account, we may set the modes $a_r^a$ and $h_{\mu r}$ for $\mu\in\{t,\vec{x},r\}$ to zero. This leads to eight constraints in addition to the equations of motion for the 22 dynamical fields. This reduces the number of the physical modes of the system to 14.

The shear viscosities are determined by the zero momentum correlation functions. Thus we only consider time dependent fluctuations, \ie $\vec{k}=0$. These fluctuations can be characterized by their transformation properties under the unbroken $SO(2)$ rotational symmetry:
\begin{equation}
\label{eq:helicitystateszeromomentum}
\begin{aligned}
&\text{helicity }2:\; &&h_{yz}, h_{yy}-h_{zz}\\
&\text{helicity }1: &&h_{yt}, h_{xy}, h_{yr}; a_y^a\\
&	&&h_{zy},h_{xz},h_{zr};a_z^a\\
&\text{helicity }0: &&h_{tt},h_{yy}+h_{zz},h_{xx},h_{xt},h_{xr},h_{tr},h_{rr}; \\
& &&a_t^a, a_x^a, a_r^a\,.
\end{aligned}
\end{equation}
The 14 physical modes decouple in several blocks. The first block contains the usual two physical helicity two modes including the shear mode $h_{yz}$. In the helicity one block there are eight physical modes which can be split in two blocks by the residual $SO(2)$ rotational symmetry. In this block the additional shear modes $h_{xy}$ and $h_{xz}$ appear. The helicity zero block contains additional four physical modes.

Since the first shear mode $h_{yz}$ is the usual helicity two mode, the general proofs of universality \cite{Kovtun:2004de,Buchel:2003tz,Benincasa:2006fu,Iqbal:2008by} apply and we obtain $\eta_{yz}/s=1/4\pi$. The second shear mode $h_{xy}$ is a helicity one mode and couples to other physical helicity one modes. Thus the assumptions of the proofs of universality \cite{Kovtun:2004de,Buchel:2003tz,Benincasa:2006fu,Iqbal:2008by} are not satisfied and the shear viscosity can be non-universal as we will see in the following.

Due to the residual $\mathbb{Z}_2$ symmetries, the four physical helicity one modes decouple  into two blocks of coupled differential equations in the case of $\vec k = 0$. The first block contains one physical mode which is determined by the dynamical fields $h_{yt}$ and $a^3_y$ and the constraint $h_{yr}=0$. This mode determines the electrical \cite{Basu:2009vv} and thermal conductivity and the thermoelectric coefficients. The three coefficients are related by Ward identities (see \eg \cite{Hartnoll:2009sz}). The second block contains the three physical modes $a^1_y$, $a^2_y$, and $h_{xy}$. We are interested in the second block which contains the shear mode  $\Psi=h_{xy}/(r^2f^2)=h_x^y$. The linearized equations of motion for the modes in the second block are given by

\begin{equation}
\begin{split}
0=&\Psi''+\left(\frac{1}{r}+\frac{4 r}{N}+\frac{6 f'}{f}-\frac{r \alpha ^2 \phi^{\prime 2}}{3 N \sigma^2}\right) \Psi^\prime+\frac{\omega^2 \Psi}{N^2 \sigma^2}\\
	&+\frac{2\alpha^2}{r^2f^2}\left(w' a^{1 \prime}_y -\frac{w \phi^2a^1_y }{N^2 \sigma^2}+\frac{i \omega w \phi a^2_y }{N^2 \sigma^2}\right)\\
0=&a^{1 \prime\prime}_y+a^{1 \prime}_y \left(\frac{1}{r}-\frac{2 f'}{f}+\frac{N'}{N}+\frac{\sigma '}{\sigma}\right)+a^1_y \left(\frac{\omega^2}{N^2 \sigma^2}+\frac{\phi^2}{N^2 \sigma^2}\right)-f^6 w' \Psi'\\
	&-\frac{2 \ii \omega a^2_y \phi}{N^2 \sigma^2}\\
0=&a^{2\prime\prime}_y+a^{2 \prime}_y \left(\frac{1}{r}-\frac{2 f'}{f}+\frac{N'}{N}+\frac{\sigma'}{\sigma}\right)+a^2_y \left(-\frac{f^4 w^2}{r^2 N}+\frac{\omega^2}{N^2 \sigma^2}+\frac{\phi^2}{N^2 \sigma^2}\right)\\
	&+\frac{\ii \omega \phi}{N^2\sigma^2}\left(-f^6 w \Psi+2 a^1_y\right)\,,
\end{split}
\end{equation}

where the prime denotes the derivative with respect to the radial variable $r$. We solve the coupled differential equations numerically and determine the retarded correlator $G^R_{xy,xy}$ using the recipe of \cite{Son:2002sd,Herzog:2002pc,Kaminski:2009dh},
\begin{equation}
 G^R_{xy,xy}(\omega,0) = +\frac{1}{2 \kappa_5^2}r^5 \frac{\Psi'}{\Psi} \bigg|_{r\rightarrow \infty} - \text{counter terms}\,.
\end{equation}
Using the Kubo formulae \eqref{eq:Kubo}, we numerically  determine the ratio of shear viscosity $\eta_{xy}$ to entropy density $s$, which is given by the Bekenstein-Hawking entropy of the black hole $s=2\pi r_h^3/\kappa_5^2$.

\begin{figure}[h]
\centering
\includegraphics[width=0.9\linewidth]{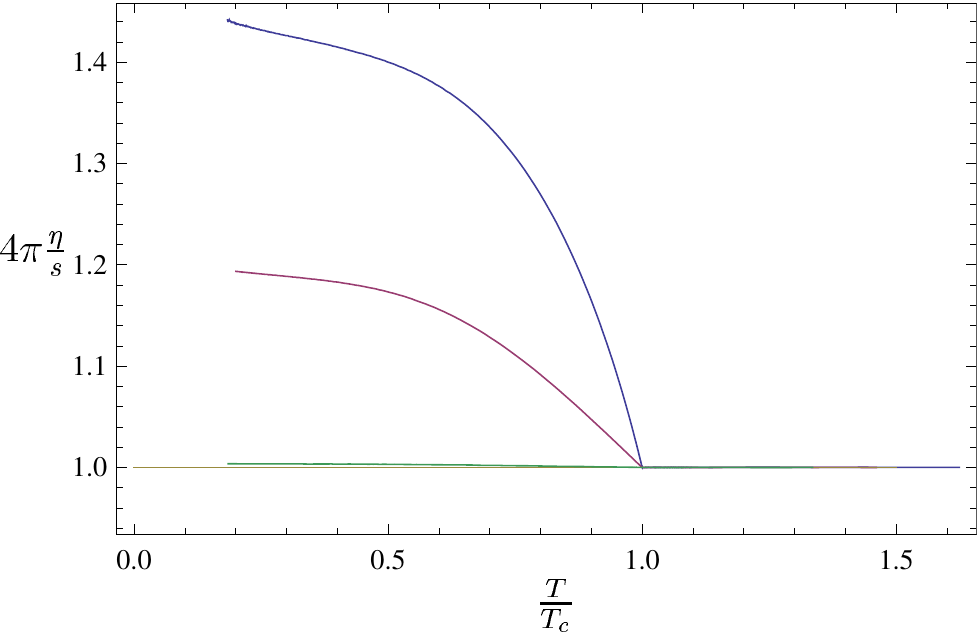}
\caption{Ratio of shear viscosities $\eta_{yz}$ and $\eta_{xy}$ to
  entropy density $s$ over the reduced temperature $T/T_c$ for
  different values of the ratio of the gravitational coupling constant
  to the Yang-Mills coupling constant $\alpha$: The color coding is as
  follows: In yellow, $\eta_{yz}/s$ for all values of 
$\alpha$; while the curve for $\eta_{xy}/s$ is plotted in green for $\alpha=0.032$, red for $\alpha=0.224$ and blue for $\alpha=0.316$. The shear viscosities coincide and are universal in the normal phase $T\ge T_c$. However in the superfluid phase $T<T_c$, the shear viscosity $\eta_{yz}$ has the usual universal behavior while the shear viscosity $\eta_{xy}$ is non-universal.}
\label{fig:viscosity}
\end{figure}

In fig~\ref{fig:viscosity} we compare our numerical results for the
ratio of the shear viscosity $\eta_{xy}$ to the entropy density $s$
with the universal behavior of the shear viscosity $\eta_{yz}$ for
different values of the ratio of the gravitational coupling to the
Yang-Mills coupling, denoted by $\alpha$. We see that in the normal
phase $T\ge T_c$, the two shear viscosities coincide as required in an
isotropic fluid. In addition, the ratio of shear viscosity to entropy
density is universal. In the superfluid phase $T<T_c$, the two shear
viscosities deviate from each other and $\eta_{xy}$ is
non-universal. However it is exciting that the KSS bound on the ratio
of shear viscosity to entropy density \cite{Kovtun:2004de} is still
valid.

\begin{figure}[h]
\centering
\includegraphics[width=0.9\linewidth]{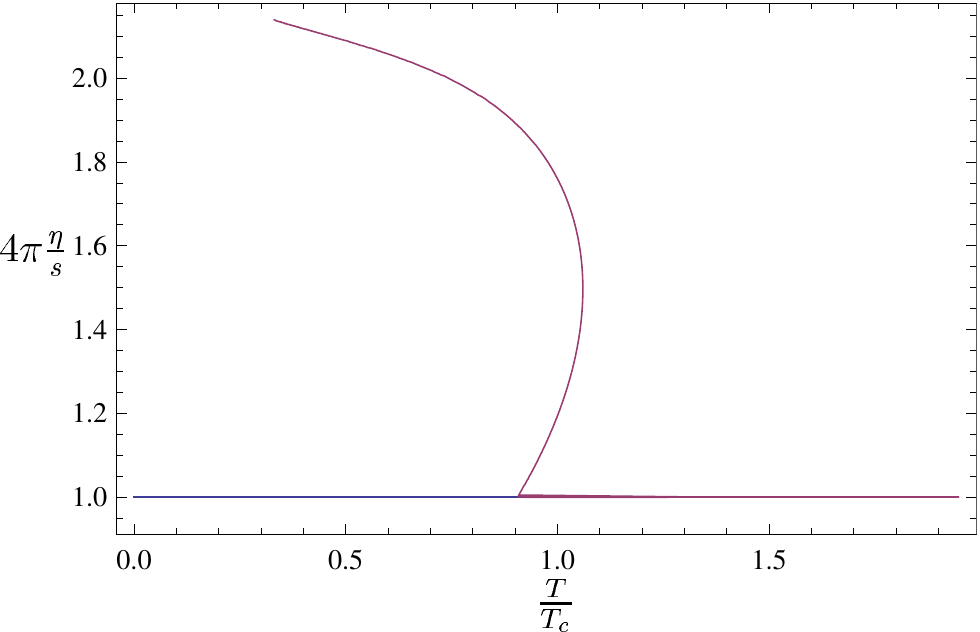}
\caption{Ratio of shear viscosities $\eta_{yz}$ (blue) and $\eta_{xy}$ (red) to entropy density $s$ over the reduced temperature $T/T_c$ for $\alpha=0.447$, which is bigger than the critical value where the phase transition becomes first order:  The shear viscosities coincide in the normal phase $T\ge T_c$ and are universal. In the superfluid phase $\eta_{xy}$ is non-universal. Close to the phase transition, it is multivalued as expected  for a first order phase transition.}
\label{fig:viscosityfirstorder}
\end{figure}

The difference between the two viscosities in the superfluid
phase is controlled by $\alpha$. In the probe limit where $\alpha=0$,
the shear viscosities also coincide in the superfluid phase. By
increasing the back-reaction of the gauge fields, \ie rising $\alpha$,
the deviation between the shear viscosities becomes bigger in the
superfluid phase as shown in fig~\ref{fig:viscosity}. If $\alpha$ is
larger than the critical value found in \cite{Ammon:2009xh} where the
phase transition to the superfluid phase becomes first order, the shear viscosities are also multivalued close to the phase transition as seen in fig.~\ref{fig:viscosityfirstorder}.  Since there is a maximal $\alpha$ denoted by $\alpha_\text{max}$ for which the superfluid phase exists, we expect that the deviation of the shear viscosity $\eta_{xy}$ from its universal value is maximal for this $\alpha_\text{max}$. Unfortunately numerical calculations for large values of $\alpha$ are very challenging such that we cannot present satisfying numerical data for this region. It is interesting that also the deviations due to $\lambda$ and $N_c$ corrections are bounded. In this case the bound is determined by causality \cite{Buchel:2009tt}.

For $\alpha$ smaller than the critical value where the phase
transition is second order, we may study the critical behavior of the
ratio of the shear viscosities to entropy density close to the phase
transition. Due to universality, $\eta_{yz}/s$ is constant and does
not change on both sides of the phase transition, while $\eta_{xy}/s$
is only constant in the normal phase, but has a different critical
behavior in the superfluid phase. Let us consider 
the critical exponent related to $\eta_{xy}/s$ and its dependence 
on $\alpha$. From our numerical data we obtain the critical behavior
\begin{equation}
\label{eq:criticalbehavior}
1-4\pi\frac{\eta_{xy}}{s}\propto \left(1-\frac{T}{T_c}\right)^\beta\quad\text{with}\quad \beta=1.00\pm 3\%
\end{equation}
in the superfluid phase close to the phase transition. It is
interesting that the critical exponent $\beta$ does not change with
$\alpha$. A more precise statement can be made if it is possible to
determine this critical exponent analytically. We suggest that this
can be achieved by an expansion in the
order parameter $\langle J_1^x\rangle$ and in the back-reaction
controlled by $\alpha$. So far, the expansion is known in the
probe limit $\alpha=0$ only \cite{Herzog:2009ci}.  However
we expect that this expansion may be extended to small values of $\alpha$. We plan to study the critical behavior of this system in more detail in the future and to make contact with the theory of dynamical critical phenomena \cite{Hohenberg:1977ym}.

To our knowledge this is the  first non-universal behavior of the shear viscosity to entropy density ratio in a theory which is holographically dual to classical, two derivative Einstein gravity theory. This non-universal behavior is due to the fact that the corresponding gravity fluctuation $h_{xy}$ transforms as a helicity one state and therefore couples to the fluctuations of the gauge field $a_y^{1,2}$. We expect that similar results can be obtained for every gravity background which spontaneously breaks the rotational symmetry.

\appendix

\section{General Remarks on viscosity in anisotropic fluids}
\label{viscosity-anisotropic}
Here we explain why the Kubo formulae \eqref{eq:Kubo} are appropriate for calculating the shear viscosities $\eta_{xy}(=\eta_{xz})$ and $\eta_{yz}$ even in a p-wave superfluid. 

In general, viscosity refers to the dissipation of energy due to any
internal motion \cite{Landau:fluid}. For an internal motion which
describes a general translation or a general rotation, the dissipation
is zero. Thus the dissipation depends  on the gradient of the
velocities $u^\mu$ only  in the combination $u_{\mu\nu}=\frac{1}{2}\left(\del_\mu u_\nu+\del_\nu u_\mu\right)$, and we may define a general dissipation function $\Xi=\frac{1}{2}\eta^{\mu\nu\lambda\rho}u_{\mu\nu}u_{\lambda\rho}$, where $\eta^{\mu\nu\lambda\rho}$ defines the viscosity tensor \cite{Landau:1959te}. Its symmetries are given by
\begin{equation}
\label{eq:symeta}
\eta^{\mu\nu\lambda\rho}=\eta^{\nu\mu\lambda\rho}=\eta^{\mu\nu\rho\lambda}=\eta^{\lambda\rho\mu\nu}\,.
\end{equation}
The part of the stress tensor which is dissipative 
due to viscosity is defined by
\begin{equation}
\label{eq:stress}
T_\text{diss}^{\mu\nu}=-\frac{\del \Xi}{\del u_{\mu\nu}}=-\eta^{\mu\nu\lambda\rho}u_{\lambda\rho}\,.
\end{equation}
We consider a fluid in the rest frame of the normal fluid $u^t=1$. To
satisfy the condition of the Landau frame $u_\mu
T_\text{diss}^{\mu\nu}=0$, the stress-energy tensor and thus the
viscosity has non-zero components only in the spatial directions $i,j=\{x,y,z\}$. In general only 21 independent components of $\eta_{ijkl}$ appear in the expressions above.

For an isotropic fluid, there are only two independent components which are usually parameterized by the shear viscosity $\eta$ and the bulk viscosity $\zeta$. The dissipative part of the stress tensor becomes $T^\text{diss}_{ij}=-2\eta(u_{ij}-\frac{1}{3}\delta_{ij}u_{l}^l)-\zeta u_{l}^l\delta_{ij}$ which is the well-known result.

In a transversely isotropic fluid, there are five independent components of the tensor $\eta_{ijkl}$. For concreteness we choose the symmetry axis to be along the $x$-axis. The non-zero components are given by
\begin{equation}
\label{eq:etatensorpwave}
\begin{aligned}
&\eta_{xxxx}=\zeta_x-2\lambda\,,
&&\eta_{yyyy}=\eta_{zzzz}=\zeta_y-\frac{\lambda}{2}+\eta_{yz}\,,\\
&\eta_{xxyy}=\eta_{xxzz}=\lambda\,,
&&\eta_{yyzz}=\zeta_y-\frac{\lambda}{2}-\eta_{yz}\,,\\
&\eta_{yzyz}=\eta_{yz}\,,
&&\eta_{xyxy}=\eta_{xzxz}=\eta_{xy}\,.
\end{aligned}
\end{equation}
The non-zero off-diagonal components of the stress tensor are given by
\begin{equation}
\label{eq:off-diagonalstressanisotroptic}
\begin{split}
&T^\text{diss}_{xy}=-2\eta_{xy} u_{xy}\,,\quad T^\text{diss}_{xz}=-2\eta_{xy} u_{xz}\,,\\
&T^\text{diss}_{yz}=-2\eta_{yz} u_{yz}\,.
\end{split}
\end{equation}

So far, we only considered the contribution to the stress tensor due
to the dissipation via viscosity and found the terms in the
constitutive equation which contain the velocity of the normal fluid
$u_\mu$. In general, also terms depending on the derivative of
Nambu-Goldstone boson fields $v_\mu=\del_\mu \varphi$, on the
superfluid velocity and on the velocity of the director may contribute
to the dissipative part of the stress tensor. Here the director is given by the vector pointing in the preferred direction.   However these terms do not contribute to the off-diagonal components of the energy-momentum tensor for the following reasons: (1) a shear viscosity due to the superfluid velocity leads to a non-positive divergence of the entropy current \cite{Landau:fluid,Pujol:2002na} and (2)  no rank two tensor can be formed out of degrees of freedom of the director if the gradients of the director vanish \cite{LESLIE01011966}. In our case the second argument is fulfilled since the condensate is homogenous and the fluctuations depend only on time. These degrees of freedom will generate additional transport coefficients, but they do not change the shear viscosities which we study in this letter. Thus we can write Kubo formulae which determine the shear viscosities in terms of the stress energy correlation functions as in \eqref{eq:Kubo}.

\paragraph{ACKNOWLEDGEMENTS}
\addcontentsline{toc}{section}{Acknowledgments}
We are grateful to A. Yarom, M. Kaminski, K. Landsteiner and especially to A. Buchel and P. Kovtun for discussions. This work was supported in part by  {\it The Cluster of Excellence for Fundamental Physics - Origin and Structure of the Universe}. PK would like to thank the Erwin Schr\"odinger Institute, Princeton University, MIT, Perimeter Institute, University of British Columbia, University of Victoria, University of Washington and University of Crete for hospitality, where part of this work was done.


\providecommand{\href}[2]{#2}\begingroup\raggedright\endgroup

\end{document}